 \documentstyle[prl,aps]{revtex}
\input epsf

\begin{document}

\preprint{12-Feb-96 RL DRAFT-1}

\draft

\twocolumn[

\title{PERFECT QUANTUM ERROR CORRECTION CODE}

\author{Raymond Laflamme$^1$, Cesar Miquel$^{1,2}$, 
Juan Pablo Paz$^{1,2}$ and Wojciech Hubert Zurek$^1$}
 
\address{\vspace*{1.2ex}
 	\hspace*{0.5ex}{$^1$Theoretical Astrophysics, T-6, MS B288}\\
 	 Los Alamos National Laboratory, Los Alamos, NM 87545, USA \\[1.2ex] 
 	 $^2$Departamento de F\'\i sica, FCEyN, Pabell\'on 1, 
         Ciudad Universitaria, 1428 Buenos Aires, Argentina}

\date{\today}
\maketitle

]


\begin{abstract}
We present a quantum error correction code which 
protects a qubit of information against general one qubit errors which maybe caused by the interaction with the environment. 
To accomplish this, we encode the  original state by distributing
quantum information over five qubits,
the minimal number required for this task. 
We give a simple circuit which takes the initial state with
four extra qubits in the state $|0\rangle$ to the encoded state.
The circuit can be converted into a decoding one by simply running it
backward. 
Reading the extra four qubits at the decoder's
output we learn which one of the sixteen alternatives (no error plus all fifteen
possible 1-bit errors)  was realized.  The original state of the encoded qubit can then be restored by a simple unitary transformation. 
\end{abstract}





Quantum computation - which has attracted so much attention as a result of progress in designing efficient quantum algorithms\cite{review,shor94} - is still far from practical implementation.  The biggest difficulty is the fragility
of the quantum states required to process information.  All the  proposed implementations  will suffer from the interaction
with the environment, and even a weak coupling may result in decoherence\cite{zurek91,unruh95,clsz95}.  Moreover, other sources of errors
(i.e., timing of laser pulses in the linear trap computer of ref.\cite{cirac95})
will add to the problem.

In classical computers, errors can also occur and are handled  through various error correcting techniques\cite{macsloane}. 
However, in the quantum case different error correction techniques are needed
to  protect quantum superposition and entanglement (which are essential ingredients of quantum computation).
The simplest scheme \cite{zurek84} of this sort can be based on a purely quantum watchdog effect.  It has been recently demonstrated to show promise\cite{miquel96},
but it suffers from an imperfection of being essentially probabilistic-- i.e.,
in principle only some of the correctable errors will actually be corrected by its application.  Thus in the terminology of the error correction community, this is scheme is not perfect\cite{macsloane}.

Shor\cite{shor95} has championed a different strategy (based on classical schemes using redundancy).   
The idea is to store quantum information not in a single qubit but in an entanglement of nine qubits.  This
scheme allows one to correct for any error incurred by any one of the nine qubits.  Steane\cite{steane} and Calderbank and Shor\cite{calde}  have proposed a different scheme which uses only seven bits for this purpose and demonstrated that this is the least required for the strategies inspired by the classical coding theory which is based on linear codes\cite{steane}.  However these codes are
not perfect as they use more bits than is absolutely necessary to correct 1-bit errors\cite{macsloane}.

In the quantum case at hand,  classical coding theory seems to be too restrictive.  All classical
codes are based on the Hamming distance \cite{hamming} (the number of different bits between  two codewords).  
Efficient quantum codes will have to use a quantum analog of this distance.  Below we
present  a perfect (i.e. capable of correcting all 1-bit errors with the minimum number of extra qubits) quantum error correction code using only 
five qubits (shown to be the smallest possible number). 
Our code is {\it not} a classical linear code \cite{steane} but a truly 
quantum code. Some of its mathematical properties are discussed
below but others certainly deserve further study. 
A notable property of our error correction code is 
that the encoding can be done using a remarkably simple circuit which
is itself the central piece of the error correction scheme enabling
us to recover from general one bit errors. 

Before presenting our perfect code, let us mention 
what are the requirements it must satisfy. An encoding of 
one qubit into $n$ qubits 
is a representation of the logical states $|0_L\rangle$ and $|1_L\rangle$  
as entangled states in the n--particle Hilbert space;
\begin{equation}
|0_L\rangle=\sum_{i=0}^{2^n-1} \mu_i |i\rangle \ ; \
|1_L\rangle=\sum_{i=0}^{2^n-1} \nu_i |i\rangle 
\,,
\label{eq:zeroone}
\end{equation}
where the states $|i\rangle=|i_{n-1},\ldots,i_0\rangle$ form a basis of the 
n--particle Hilbert space with $i_j$ defining the binary 
representation of the integer $i$. 
To serve as a quantum error correction code Eq.(\ref{eq:zeroone})
must satisfy certain conditions whose origin is best understood 
by analyzing the effect of the interaction with the environment. 
A general interaction between the k-th qubit and its 
environment will lead to an evolution of the form;
\begin{eqnarray}
|e\rangle |0_k\rangle \rightarrow 
|e_0\rangle|0_k\rangle  + \ |e_0^B\rangle|1_k\rangle \, \, \ \nonumber   \\  \
|e\rangle|1_k\rangle \rightarrow
|e_1\rangle |1_k\rangle  + \ |e_1^B\rangle|0_k\rangle    
\,,
\label{eq:onebitint}
\end{eqnarray}
where  $|e\rangle$, $|e_{0,1}\rangle$, $|e_{0,1}^B\rangle$ are states of the environment which 
will remain arbitrary throughout this paper 
(apart from the obvious orthogonality and normalization constraints 
imposed by unitarity of the evolution in Eq.(\ref{eq:onebitint})). 
The effect of the interaction given by Eq.(\ref{eq:onebitint}) upon the logical states
$|0_L\rangle$ and $|1_L\rangle$ is easily calculated;
\newpage

\begin{equation}
|e\rangle 
    \begin{array}{c} |0_L\rangle \\ |1_L\rangle \end{array}
  \rightarrow  
\Big (|e_+\rangle {\cal I} + |e_-\rangle \sigma_z^k + 
|e^B_+\rangle \sigma_x^k - |e^B_-\rangle i\sigma_y^k \Big )
 \begin{array}{c} |0_L\rangle \\ |1_L\rangle \end{array}
 \,,
\label{eq:kbiterr}
\end{equation}
where $\sigma_i^k$ are the Pauli matrices acting on the k-th bit.
The states of the environment appearing in Eq.(\ref{eq:kbiterr})
are $|e_\pm\rangle=(|e_0\rangle \pm |e_1\rangle)/2$ and 
$|e^B_{\pm}\rangle=(|e^B_0\rangle \pm |e^B_1\rangle)/2$.  
Four types of outcome due to 
interaction with the environment exhaust all possibilities.
First, the state may remain unchanged (the operator ${\cal I}$ is 
proportional to the  identity).  Second,
the state of the system may pick a minus sign in front of all the 
states with a 1 in the k-th qubit (thus corresponding to action of the operator $\sigma_z^k$).  This alternative is  correlated with 
the environment $|e_-\rangle$. 
Third, the state of the system 
may be altered by flipping the k-th bit (through the operator $\sigma_x^k$)
getting correlated with the states $|e^B_+\rangle$. 
Fourth, and finally, the system may get a 
bit flip in the k-th bit together with a sign flip for which the operator is
$-i\sigma_y^k$, an option correlated with $|e^B_-\rangle$.
The second operation is denoted by $S_k$ (for {\it sign} flip), the third by
$B_k$ (for {\it bit} flip) and the fourth one by $BS_k$ (which is self-explanatory). 
Note that the same state of the environment is coupled to the respective states
of $|0_L\rangle$ and $|1_L\rangle$.  This is essential in what follows.

The defining property of a quantum error correction code 
Eq.(\ref{eq:zeroone}) is the following: the original two dimensional Hilbert space spanned by $|0_L\rangle$ and $|1_L\rangle$ 
must be mapped coherently into orthogonal 2-dimensional  Hilbert spaces  
corresponding to each of the different environment--induced errors 
(denoted as $S_k$, $B_k$ and $BS_{k}$). 
This is sufficient to recover from a 1-qubit error  since 
it is possible to measure in which 2-d Hilbert space the system is
without destroying the relevant coherence.
After the 
measurement it is possible to restore the original quantum 
state by means of simple unitary transformations (which depend upon 
the result of the measurement). 

Orthogonality of the subspaces corresponding to the different 
errors  imposes a rather stringent constraint on the dimension of the 
Hilbert space which must be large enough to accommodate so many orthogonal 
subspaces. How big should this space be? 
Orthogonality requires a subspace for each of the
three errors every qubit can suffer and another one for the unperturbed 
logical state. This makes a total of $3n+1$. We must double this
to have enough space to accommodate both logical states 
and their erroneous descendants. Thus, the number of subspaces 
is $2(3n+1)$. To have enough room in the Hilbert space the condition;
\begin{equation}
2(3n+1)\le 2^n 
\,,
\label{eq:dimens}
\end{equation}
must be satisfied. 
Both Shor's $n=9$--code and Steane's $n=7$--code satisfy
this constrain while $n=5$ is the smallest number which saturates
Eq.(\ref{eq:dimens}). The code we present has 5 bits.  

The orthogonality conditions can be written as algebraic constraints on
the coefficients $\mu_i$ and $\nu_i$ which define the encoding. For the
sake of space and time we will not write them all explicitly but just mention
the following simple subset;
\begin{equation}
\sum_{k-even\atop{l-even}}|\mu_i|^2=
\sum_{k-even\atop{l-odd}}|\mu_i|^2=
\sum_{k-odd\atop{l-even}}|\mu_i|^2=
\sum_{k-odd\atop{l-odd}}|\mu_i|^2
\,,
\label{eq:sigmacond}
\end{equation}
for all $k,l=1,\dots,5$ (and a similar condition for $\nu_i$). 
The sums are over k--even and k--odd numbers: k--even (k--odd) 
numbers are those with a 0 (1) 
in the k-th bit.  If we restrict ourselves to encodings
satisfying $|\mu_i|=|\nu_i|=1$, an assumption based on simplicity, 
the above condition implies that we need at least eight states in the 
superposition. Thus, five bits and eight states in the superposition 
seem to be the minimum required by the orthogonality conditions
(and the simplicity assumption). Moreover, it is easily shown that 
it is impossible to satisfy all the constraints by using only positive 
numbers for $\mu$s or $\nu$s ($+1$ in our case) so either phases or minus signs are essential.  

The conditions of Eq.(\ref{eq:sigmacond}), while still incomplete, are nevertheless extremely restrictive: In fact, one can 
prove that they essentially determine (up to permutations between 
bits) what are the eight states $|i\rangle$ allowed in the 
superposition of Eq.(\ref{eq:zeroone}).  This determines the encoding of each of the  
logical states, thus defining the support of 
the code.  It is interesting to note that the solution can be guessed from
Steane's encoding \cite{steane} by dropping any two of its qubits.
The only remaining freedom is in the sign 
distribution between states, which can be found by a computer search.
This is how we have first arrived at the class of possible encodings
exemplified by the  following perfect $5$--bit code
\begin{eqnarray}
|0_L\rangle =\ \ \, |b_1\rangle|00\rangle-|b_3\rangle|11\rangle+
|b_7\rangle|10\rangle+|b_5\rangle|01\rangle \nonumber \ \ \\
 |1_L\rangle=-|b_2\rangle|11\rangle-|b_4\rangle|00\rangle+
|b_8\rangle|01\rangle-|b_6\rangle|10\rangle 
\,,
\label{eq:thecode2}
\end{eqnarray}
where the (unnormalized) 3--particle Bell states are defined 
as $|b_{1\atop 2}\rangle=(|000\rangle\pm|111\rangle)$,
$|b_{3\atop 4}\rangle=(|100\rangle\pm|011\rangle)$, $|b_{5\atop 6}\rangle=(|010\rangle\pm|101\rangle)$,
$|b_{7\atop 8}\rangle=(|110\rangle\pm|001\rangle)$. 
Other allowed codes can be found from Eq.(\ref{eq:thecode2}) by permutations
of bits and coordinated  sign changes. Thus, all 
the allowed codes have the same sign pattern, with two minus signs
in one of the logical states and four in the other (these 
results will be proven in detail elsewhere). 
The mathematical
structure behind this sign distribution (which,
as we said before, is the only freedom we have, save for the `gauge transformation' in the form of sign and coordinated bit flips) still lies beyond
our present understanding.  

The encoding Eq.(\ref{eq:thecode2}) can be implemented by using the 
circuit depicted in Figure 1a. The original information carrier is
the qubit $|Q\rangle$ which may be in a general state $|Q\rangle=\alpha|0\rangle+\beta
|1\rangle$. After the action of the encoding circuit, and when the other 
input states are all set to $|0\rangle$, the output state will always
be given by $\alpha|0_L\rangle+\beta|1_L\rangle$. This circuit is just
a combination of quantum logic gates (controlled--not, controlled
rotations, etc.) which can be implemented (at least {\it in 
principle}) in various physical settings. 

Until now we exhibited a quantum code and a quantum circuit 
which acts as encoder. However, the error correction method would
not be complete without the circuit for actually {\it correcting} 
all the possible one bit errors. The most remarkable feature of our 
method is that the circuit for this is {\it exactly the same 
as the one for encoding but run backwards} (see Figure 1b). This is in contrast 
with all previous schemes discussed in the literature where a different
decoding/correction circuit was necessary. 

A heuristic argument has guided us in searching for this 
circuit. The fact that we are using exactly $n=5$ bits allows 
us {\it in principle} to have a circuit such us the one we found. 
For, to distinguish the $16$ different error syndromes (the ``no
error alternative'' plus the $15$ ones corresponding to five errors
of each type $S_k$, $B_k$ and $BS_k$) we would need to make four binary 
tests (which would provide us with 16 results). This is precisely
what the circuit does: when any one of the sixteen possible states
inputs the encoder from the right, the states $|a'\rangle$, $|b'\rangle$, $|c'\rangle$ 
and $|d'\rangle$ uniquely identify the input and allow us to know what 
the state of the qubit $|Q'\rangle$ is. All possibilities are exhibited 
in Table 1. Some of them are easily understood. For example, the trivial  
case $|a'\rangle=|b'\rangle=|c'\rangle=|d'\rangle=|0\rangle$ corresponds to the ``no error'' alternative 
(since in that case the input in the left is identical to the 
one used for encoding).
Other alternatives, such as the one corresponding to the $S_1$ 
syndrome (an error
in the first bit) can be easily identified by looking at the 
circuit from the left to the right: In fact, if the input to the 
encoder is not $|a\rangle=|b\rangle=|c\rangle=|d\rangle=|0\rangle$ but $|a\rangle=|1\rangle$,
$|b\rangle=|c\rangle=|d\rangle=|0\rangle$ the output state is easily seen to be the one
corresponding to the $S_1$ error (since the first rotation would 
produce a state with a minus sign in front of the $|1\rangle$ state). 
Other alternatives are less obvious but they all work in the same
way. 

Thus, after using the encoding circuit in backwards direction  we have a 
precise diagnosis of what went wrong (if anything) with our quantum 
bit. The state of the qubit $|Q'\rangle$ may be easily restored to 
the original $\alpha|0\rangle+\beta |1\rangle$ by a unitary transformation 
which depends upon the measurement of the states $|a'\rangle,|b'\rangle,|c'\rangle$ and $|d'\rangle$\cite{quick}.

Assuming that the interaction affected at most one bit in any way, we have shown
that there exist a 5-qubit code which corrects perfectly, 
i.e. has perfect fidelity
\cite{schu95}.  It is not difficult to convince yourself that 
if the probability of an error in only one qubit is $p$, the fidelity
of the code where the restriction to only one error is lifted will
be $1-cp^2+\dots$, for some constant $c$.    This is an improvement
on the uncorrected evolution of a single qubit which has fidelity $1-p$ 
as long as $c<p$.

The support of our code is unique under the conditions; $i$) that the  
coefficients of the codewords have unit modulus, and;
$ii$) that under error due to the interaction with the environment 
the logical states would go to mutually orthogonal states\cite{manlaf}.

{\bf Acknowledgements.}
We would like to thank E.  Knill for many useful comments concerning classical and quantum errors correction codes and A. Ekert as well as R. Hughes for general comments about quantum computation.

%

{\ }
\epsfxsize=3.5truein\epsfbox{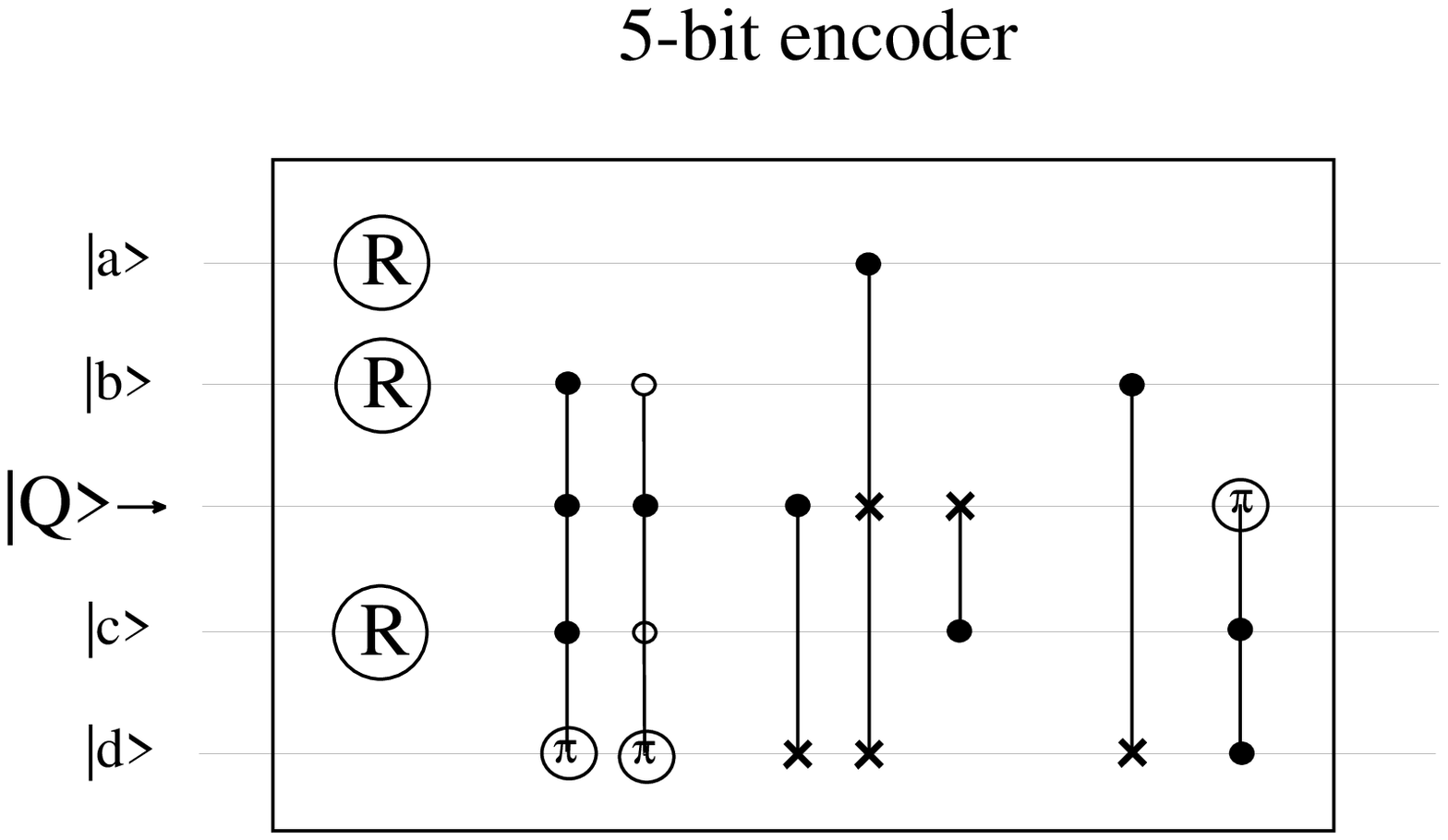}

Figure 1a.  Circuit for the encoding of the states described in Eq.(6).  R describes
the rotation $|0\rangle\rightarrow (|0\rangle + |1\rangle)/\sqrt{2}$ and
 $|1\rangle\rightarrow (|0\rangle - |1\rangle)/\sqrt{2}$ . The element
with an $\times$ corresponds to a control not (with control on the filled 276circle);
if the control is $|1\rangle$ then the state at $\times$ is flipped.
The element including $\pi$ correspond to a conditional rotation
by a phase $\pi$, where the condition is satisfied  
when the state has the bit in the 0 state for the empty circle
and 1 for the filled one.

\hskip 1.7 truein\vbox{\tabskip=0pt \offinterlineskip
\def\tablerule{\noalign{\hrule}}
\halign to 175 pt
     {\strut#&\vrule#\tabskip=0pt& \vrule # &\vrule
      \hfil #  \vrule  \tabskip=0pt\cr\tablerule
& & \hfil    &  \cr
&\ \ Error &  \ \ \ \ Syndrome  &\ \ Resulting state  \cr
& & \hfil $|a'b'c'd'\rangle$\hfil    & \hfil $|Q^{'}\rangle$ \hfil\cr\tablerule
&\hfil None& \hfil 0000   \hfil & $\alpha |0\rangle+\beta|1\rangle$ 
      \cr\tablerule 
&\hfil BS3\hfil& \hfil 1101   \hfil & $-\alpha |1\rangle+\beta|0\rangle$ 
      \cr\tablerule 
&\hfil BS5\hfil& \hfil 1111   \hfil & $-\alpha |0\rangle+\beta|1\rangle$ 
      \cr\tablerule 
&\hfil B2\hfil& \hfil 0001   \hfil &   \cr
&\hfil S3\hfil& \hfil 1010    \hfil& $\alpha |0\rangle-\beta|1\rangle$  \cr
&\hfil S5\hfil& \hfil 1100    \hfil&   \cr
&\hfil BS2\hfil& \hfil 0101    \hfil&   \cr \tablerule
&\hfil B5\hfil& \hfil 0011   \hfil &   \cr
&\hfil S1\hfil& \hfil 1000   \hfil & $-\alpha |0\rangle-\beta|1\rangle$  \cr
&\hfil S2\hfil& \hfil 0100   \hfil &   \cr
&\hfil S4\hfil& \hfil 0010    \hfil&   \cr \tablerule
&\hfil B1\hfil& \hfil 0110   \hfil &   \cr
&\hfil B3\hfil& \hfil 0111   \hfil &   \cr
&\hfil B4\hfil& \hfil 1011   \hfil & $-\alpha |1\rangle-\beta|0\rangle$  \cr
&\hfil BS1\hfil& \hfil 1110   \hfil &   \cr
&\hfil BS4\hfil& \hfil 1001   \hfil &   \cr \tablerule
\cr}}

\noindent Table 1.
Error with corresponding syndromes and states for the decoder shown in Figure
1. $B,\ S,\ BS$ correspond to a bit, a sign, or a bit and a sign flipped with the following
number which identifies the bit. To recover the initial state, 5 different unitary operations must be performed consisting of bit and sign flips on the state $|Q^{'}\rangle$.

\newpage

{\ }
\vskip -0.155truein
\epsfxsize=3.66truein\epsfbox{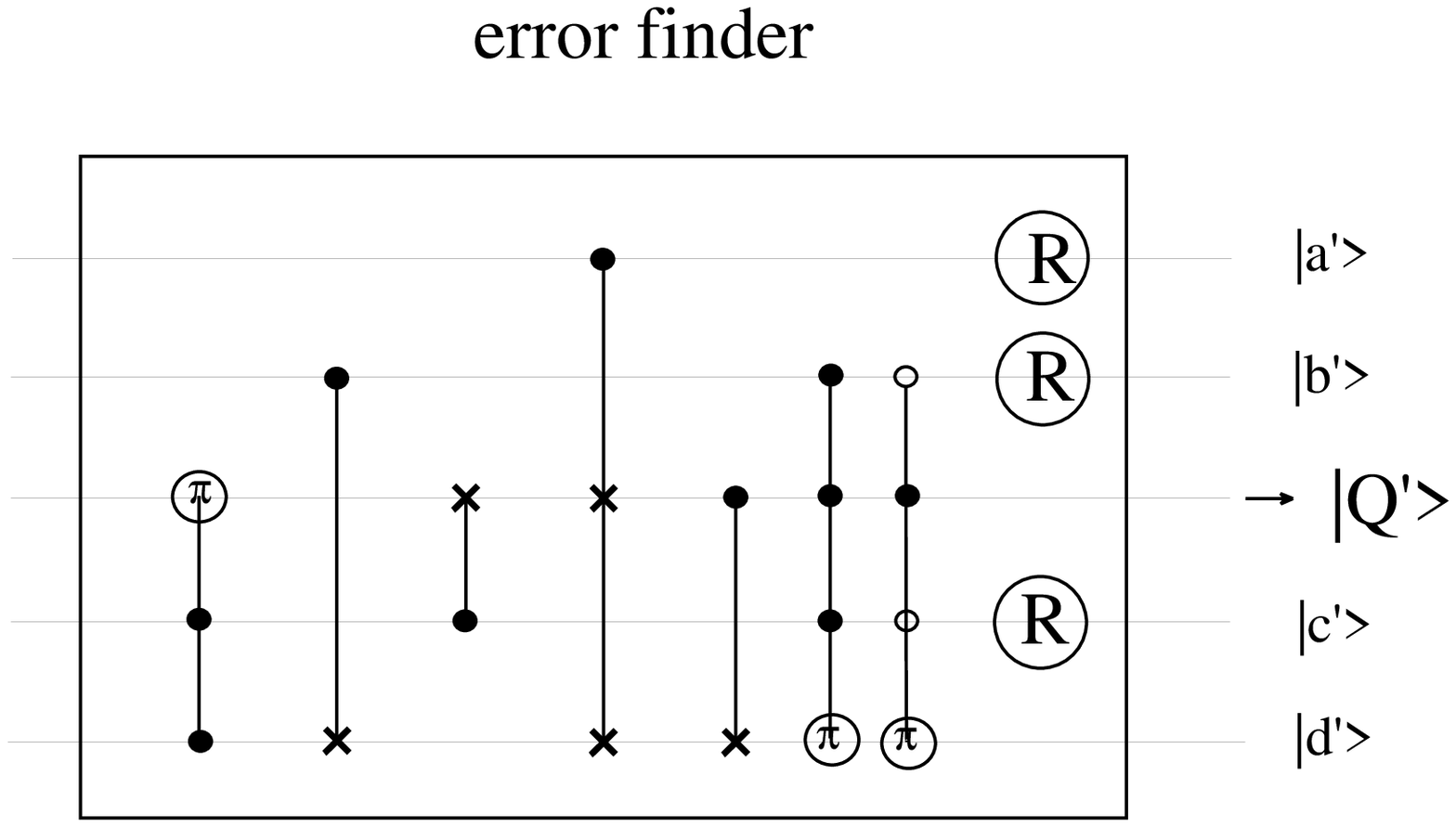}

Figure 1b.  Circuit of Figure 1a 
ran in the opposite way. 
The state $|a',b',c',d'\rangle$ gives the syndromes of table 1.
A unitary transformation brings back $|Q^{'}\rangle$ to $|Q\rangle$, which can be reencoded
using the circuit of Figure 1a.
 

\begin{thebibliography}{99}
\bibliographystyle{prsty}  
\bibitem{review}  For reviews see
S. Lloyd, {\em Scientific American} 273, p. 140, October 1995.
A. Ekert and R. Jozsa, Notes on Shor's efficient algorithm for
factoring on a quantum computer, {\em Rev. Mod. Phys.}, in press .  
 
\bibitem{shor94}
P.~W.~Shor.
\newblock In {\em Proc. 35th Annual Symposium on Foundations of Computer
   Science}, ed. S. Goldwasser. (IEEE Computer Society Press, Nov. 1994)
   pp. 124--134.

\bibitem{zurek91}
W.~H. Zurek.
\newblock {\em Physics Today}, 44:36, 1991.

\bibitem{unruh95}
W.~G. Unruh.
\newblock {\em Phys.Rev. A}51, 992, 1995.

\bibitem{clsz95}
\newblock I.Chuang, R.Laflamme, P.Shor and W.H.Zurek; {\em Science} 270, p.1633, 1995.

\bibitem{cirac95}
J.~I.~Cirac and P.~Zoller.
\newblock {\em Phys. Rev. Lett.}, 74:4091, 1995.

\bibitem{macsloane}
\newblock
F.J. MacWilliams and N.J.. Sloane, The theory of Error-Correcting Codes,
North-Holland Publishing Company, New-York, 1977.


\bibitem{zurek84}
W.~H. Zurek.
\newblock {\em Phys. Rev. Lett.}, 53:391, 1984.

\bibitem{miquel96}
\newblock C. Miquel, J.P. Paz and R. Perazzo, preprint quant-ph/9601021

\bibitem{shor95}
\newblock  P.W. Shor {\it Phys.Rev. A} 52, p.2493, 1995.

\bibitem{steane}
\newblock  A. Steane, Multiple particle interference and quantum error correction, preprint quant-ph/9601029, to be published in Proc.Roy.Soc. London.

\bibitem{calde}
\newblock A.R. Calderbank and P.W. Shor, Good quantum error-correcting codes exist, preprint quant-ph/9512032.

\bibitem{hamming}
E.~A. Lee and D.~G. Messerschmitt.
\newblock {\em Digital Communication}.
\newblock Kluwer Academic Publishers, 1988.


\bibitem{schu95}
\newblock  B. Schumacher, {\it Phys.Rev. A} 51, 2738, 1995.

\bibitem{quick}
\newblock This last unitary transform can be omitted if we redefine the meaning 
logical 0 and 1 in the quantum program.

\bibitem{manlaf}
\newblock 
This last condition is sufficient for the code to operate properly but is not
necessary.  It is possible to find codes such that some errors are mapped to the same 2-d subspace.   This alternative, which is now under investigation
does not allow less than five qubits  but might allow to detect or even correct more than the 1-qubit errors
(E. Knill and R. Laflamme, The theory of quantum error correcting
codes, in preparation).


\end{thebibliography}
\end{document}